\documentclass[aps,prl,10pt,longbibliography,twocolumn,showpacs,amsmath,amssymb,superscriptaddress,floatfix,nobibnotes]{revtex4-2}

\usepackage[utf8]{inputenc}
\usepackage{comment}
\usepackage{amsfonts,amsmath,amssymb,mathtools}  
\usepackage[colorlinks=true,breaklinks=true,allcolors=blue]{hyperref}
\usepackage{orcidlink}
\usepackage{bm}
\usepackage{physics}
\usepackage{dsfont}
\usepackage{enumitem}
\usepackage{mathtools}
\usepackage{amsfonts}
\usepackage{mathrsfs}
\usepackage{physics}
\usepackage{slashed}
\usepackage{tensor}
\usepackage{dsfont}
\usepackage{bbm}
\usepackage{booktabs}
\usepackage{multirow}
\usepackage{graphicx}
\usepackage{color}
\usepackage{array}
\usepackage[abs]{overpic}

\usepackage{tikz}
\usetikzlibrary{decorations.pathmorphing}

\usepackage{placeins}

\usepackage{makecell}

\usepackage{xspace}
\usepackage{siunitx}
\usepackage{xfrac}
\usepackage{hyperref}
\usepackage[nameinlink]{cleveref}
\usepackage{appendix}

\usepackage{adjustbox}

\usepackage{xifthen}
\usepackage{xcolor}
\hypersetup{
	colorlinks,
	linkcolor={red!75!black},
	citecolor={blue!75!black},
	urlcolor={blue!75!black}
}
\usepackage[normalem]{ulem}

\newcommand{\gettitle}{False vacuum decay in long-range interacting quantum systems}

\newcommand{\getZuerichAffiliation}{\affiliation{Institut f{\"u}r Theoretische Physik, ETH Z{\"u}rich, Wolfgang-Pauli-Str. 27, 8093 Z{\"u}rich, Switzerland}}

\newcommand{\getTriesteAffiliation}{\affiliation{CNR-INO, Area Science Park, Basovizza, I-34149 Trieste, Italy}}

\newcommand{\getHarvardChemAffiliation}{\affiliation{Department of Chemistry and Chemical Biology, Harvard University, Cambridge, Massachusetts 02138, USA}}

\hypersetup{
	pdftitle={\gettitle},
	bookmarksopen=true,
	bookmarksopenlevel=2,
	bookmarksnumbered=true
}

\begin{document}
	
\title{\gettitle}

\author{Valerio Pagni \orcidlink{https://orcid.org/0009-0007-5323-4651}}
\thanks{V.P. and L.B. contributed equally to this work.\\
E-mail V.P. at: vpagni@ethz.ch\\  E-mail L.B. at: lbatini@ethz.ch.}
\getZuerichAffiliation\getHarvardChemAffiliation
\author{Laura Batini \orcidlink{https://orcid.org/0009-0007-4129-4251}}
\thanks{V.P. and L.B. contributed equally to this work.\\
E-mail V.P. at: vpagni@ethz.ch\\  E-mail L.B. at: lbatini@ethz.ch.}
\getZuerichAffiliation
\thanks{These authors contributed equally to this work.}
\author{Nicolò Defenu \orcidlink{https://orcid.org/0000-0002-3401-3665}}\getZuerichAffiliation\getTriesteAffiliation

\newcommand{\Frac}[1]{(-\partial_x^2)^{#1}}
\newcommand{\Afrac}{\mathbb{A}}
\newcommand{\Bop}{\mathbb{B}}
\newcommand{\vF}{\bm{F}}
\newcommand{\bigO}{\mathcal{O}}  

\begin{abstract}
We formulate false-vacuum decay in a mixed-field Ising chain with $r^{-(1+\sigma)}$ interactions as a spatially nonlocal Euclidean $\phi^4$ theory featuring a fractional spatial kinetic term $\sim |q|^\sigma$. The nonlocal bounce is anisotropic in space-time and develops algebraic spatial tails. Combining thin-wall arguments with numerical solutions of the full nonlocal saddle, we find two nucleation regimes separated by $\sigma = 1$. For $0<\sigma<1$, the lifetime exponent
scales with the energy bias $h$ of the metastable state as $B\sim h^{-1/\sigma}$; for $1<\sigma<2$, the leading Coleman scaling $B\sim h^{-1}$ is recovered, while long-range effects are retained in the subleading correction $\sim h^{\sigma-2}$. The same exponents are recovered directly within a two-kink description of the Ising chain on the lattice. Our results show that tunable long-range interactions fundamentally reshape bubble nucleation and alter false-vacuum decay in quantum simulators.
\end{abstract}


\maketitle

\emph{Introduction}---The problem of false-vacuum decay has profoundly shaped our understanding of nonperturbative phenomena in quantum field theory. Today, false-vacuum decay and bubble nucleation have been observed in ferromagnetic superfluids~\cite{zenesini2024observation}, Rydberg-atom arrays~\cite{chao2026probing}, trapped-ion systems~\cite{luo2025quantum}, and programmable quantum annealers~\cite{vodeb2025stirring}; related protocols and simulations have addressed false-vacuum decay in quantum spin systems,
from one-dimensional chains~\cite{lagnese2021false, lagnese2024detecting,johansen2025many,darbha2024false,darbha2024long} to two-dimensional models~\cite{balducci2022localization,balducci2023interface,borla2026microscopic,pavevsic2026false}, while a general many-body theory of metastability has recently been developed~\cite{yin2025theory}. Beyond fundamental theoretical interest, this broad experimental activity is rooted in the fact that highly entangled quantum states---a fundamental resource for quantum information processing and metrology~\cite{horodecki2009quantum,cao2024multi}---naturally tend to occupy configurations that are metastable with respect to the microscopic Hamiltonian.

The route to creating and stabilizing highly entangled states in modern quantum simulators often exploits their most exotic features. Among those, long-range interactions have been shown to boost information propagation~\cite{richerme2014non,jurcevic2014quasiparticle} and provide a natural source for novel metastable phenomena~\cite{defenu2021metastability,defenu2024out}. Power-law couplings, $\sim r_{ij}^{-\alpha}$, are routinely engineered in trapped ions, Rydberg arrays, dipolar gases, and cavity-mediated systems~\cite{britton2012engineered,islam2013emergence,zhang2017observation53qubit,zeiher2017coherent,borish2020transverse,pagano2020quantum,li2023probing,lahaye2009physics,ritsch2013cold,periwal2021programmable,defenu2023long}, where the interaction range can be tuned from nearly all-to-all to effectively short-ranged. This tunability modifies correlation spreading, entanglement growth, collective modes, and relaxation dynamics~\cite{schachenmayer2013entanglement,hastings2006spectral,foss2015nearly,tran2019locality,lerose2020origin,defenu2021metastability,defenu2023long,defenu2024out}. Understanding how metastable states of long-range interacting quantum systems ultimately decay is therefore both a fundamental question and a practical one for quantum simulators, where long-lived nonequilibrium states are often central resources.

In local systems the metastable vacuum decays not uniformly, but via the rare nucleation of a critical droplet of stable phase, whose subsequent growth consumes the false-vacuum background.
Semiclassically, the quantum nucleation of the critical droplet is governed by the Coleman bounce, a finite-action saddle-point configuration of the Euclidean theory~\cite{coleman1977fate,callan1977fate}. 
\begin{figure}
    \centering
    \includegraphics[width=.95\linewidth]{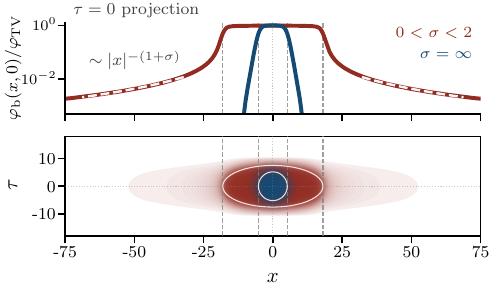}
    \caption{\textbf{The long-range critical bubble.}
Long-range interactions break the isotropy of the Coleman bounce, generating an anisotropic critical droplet with algebraic spatial tails. The lower panel shows the full bounce profile $\varphi_{\rm b}(x,\tau)$, while the upper panel ($\tau=0$ cut) contrasts the long-range bubble with the exponentially localized short-range limit. Dashed lines highlight the asymptotic algebraic tails, $\varphi_{\rm b}(x,0)\sim |x|^{-(1+\sigma)}$, which provide the real-space hallmark of nonlocal nucleation. Shown is a representative solution for $\sigma=1.1$ and $h=0.1$.}
    \label{fig:critical_bubble}
\end{figure}
Its thin-wall structure reflects locality, with a local surface tension at the droplet boundary competing against the bulk energy gain. Despite the central role of this mechanism in quantum field theory and cosmology---from early-universe phase transitions~\cite{linde1981fate} to the stability of the electroweak vacuum~\cite{isidori2001metastability,buttazzo2013investigating}---the lack of its generalization to the context of long-range quantum many-body theories shall not surprise. In fact, most existing field-theoretical descriptions retain the local Coleman logic, in which the critical droplet is governed by a short-range wall tension. For power-law interactions this assumption is no longer automatic: distant parts of the droplet interact directly, and the nucleation barrier need not be reducible to a local surface term~\cite{mccraw1980metastability,bray1993domain,hoyos2007local,mukamel2009notesstatisticalmechanicssystems,van2019nucleation,jacquier2024isoperimetric,jacquier2025exploring,lazarides2026apparent}. The lack of a comprehensive theory of false-vacuum decay in long-range \emph{quantum} many-body systems is therefore a pressing gap, limiting both the prospects for realizing robust entangled states and the understanding of structural phase stability in systems with competing local and long-range interactions~\cite{kirkby2026morphological}.

Here, we address this problem in two complementary descriptions: a spatially nonlocal field theory and a microscopic two-kink description of the long-range Ising chain.
At long wavelengths, the power-law interaction generates a fractional spatial kinetic term with $\sigma=\alpha-1$, while imaginary time remains local~\cite{defenu2017criticality,defenu2023long}, so the false-vacuum decay rate is controlled by a nonlocal Euclidean saddle point. 
As illustrated in Fig.~\ref{fig:critical_bubble}, the resulting bounce is strongly anisotropic, with algebraic spatial tails. Both descriptions yield the same lifetime scaling, while a mean-field and spin-wave matching~\cite{SuppMat} fixes the nonuniversal continuum coefficients, enabling parameter-specific comparison with quantum simulators.

Combining thin-wall arguments, numerical solutions of the full nonlocal saddle, and an independent microscopic two-kink calculation, we find two regimes separated by $\sigma=1$. The decay exponent scales as $B\sim h^{-1/\sigma}$ for $0<\sigma<1$, while for $1<\sigma<2$ the Coleman law $B\sim h^{-1}$ is recovered with a long-range correction $\sim h^{\sigma-2}$. A key result is that the thin-wall scaling laws remain valid for the full bounce despite its algebraic tails. Hence, the interaction range directly controls the bubble structure and the false-vacuum lifetime.

\emph{From spins to fields}---We start from the long-range mixed-field Ising chain
\begin{equation}
    H =
    -\sum_{i<j} \frac{J_0}{r_{ij}^{\alpha}} \sigma_i^z \sigma_j^z
    - \Gamma \sum_i \sigma_i^x
    - h_z \sum_i \sigma_i^z ,
\label{eq:Hamiltonian}
\end{equation}
where $J_0>0$ sets the ferromagnetic interaction scale, $\alpha>d=1$ controls the range of the power-law couplings, $\Gamma$ drives quantum fluctuations, and $h_z$ is a weak longitudinal field. 
For $h_z=0$ and $|\Gamma|<\Gamma_c(\alpha)$, the model has two symmetry-related ferromagnetic ground states. 
A finite $h_z$ explicitly breaks the $\mathbb{Z}_2$ symmetry and makes one of these states energetically favored (true vacuum), while the other becomes metastable (false vacuum). 
False-vacuum decay then corresponds, in spin language, to the nucleation of a droplet of the stable magnetization inside the metastable background. The transverse field $\Gamma$ supplies the quantum fluctuations allowing the system to tunnel through the energy barrier.

This model is directly relevant to trapped-ion simulators, where phonon-mediated interactions generate programmable Ising couplings~\cite{zhang2017observation53qubit,defenu2023long}. Recent experiments have observed quantum bubble nucleation in mixed-field Ising chains with engineered exponentially decaying couplings, $J_{ij}=J_0\, e^{-\beta(|i-j|-1)}$~\cite{luo2025quantum}. 
Since trapped-ion platforms natively realize power-law interactions, Eq.~\eqref{eq:Hamiltonian} is the natural long-range extension of this setting, with an intrinsically nonlocal droplet cost.

At long wavelengths, both a coarse-graining procedure and linear spin-wave theory~\cite{SuppMat} yield a local temporal kernel and the nonanalytic spatial kernel $|q|^\sigma$, with $\sigma=\alpha-1$. The resulting Euclidean action for the coarse-grained longitudinal magnetization $\phi(x,\tau)$ is
\begin{equation}
\label{eq:action_general}
S =
\int_{\tau,x}
 \Big[
\frac{c_\tau}{2}(\partial_\tau \phi)^2
+ \frac{c_{\rm LR}}{2}
\phi (-\partial_x^2)^{\frac{\sigma}{2}} \phi + V_{\rm eff}(\phi)
\Big],
\end{equation}
where here and in the following we use the shorthand
$\int_{\tau,x}\equiv \int \mathrm{d}\tau\,\mathrm{d}x$, $(-\partial_x^2)^{\frac{\sigma}{2}}$ is the fractional Laplacian~\cite{lischke2020fractional}, whose Fourier representation is $|q|^\sigma$, and $\sigma = \alpha-1$. This action belongs to the class of fractional Lifshitz field theories~\cite{basak2024massless,benedetti2024finite,benedetti2024dynamic}. We restrict to $\sigma>0$, corresponding to the additive long-range regime.
For $\sigma\le0$, the power-law interaction is nonintegrable at large distances -- where the nonanalytic $|q|^\sigma$ term generated by the power-law interaction is the leading small-momentum spatial contribution. For $\sigma>2$, the ordinary $q^2$ stiffness dominates instead, and the leading long-wavelength nucleation problem becomes effectively short-ranged.

The tilted potential $V_{\rm eff}(\phi) = \tfrac{g}{4}(\phi^2-\phi_0^2)^{2} - h \phi$ is the minimal coarse-grained description of the metastable landscape. 
For $h=0$ it has two degenerate minima at $\phi=\pm\phi_0$, while for $h\propto h_z\neq0$ one minimum is the true vacuum $\phi_{\rm TV}$ and the other the false vacuum $\phi_{\rm FV}$. 
Although the microscopic potential is generally nonpolynomial~\cite{SuppMat}, the quartic form captures the nucleation scaling whenever two locally stable minima are separated by a barrier.

The saddle point controlling the decay rate is obtained by extremizing the Euclidean action:
\begin{equation}
\label{eq:bounce_eom}
    \left[-c_\tau \partial_\tau^2
    + c_{\rm LR}(-\partial_x^2)^{\frac{\sigma}{2}}\right] \phi
    + V'_{\rm eff}(\phi)
    =0.
\end{equation}
The finite-action solution, the bounce $\phi=\phi_{\rm b}$, approaches the false vacuum at large Euclidean distance,
$  \phi_{\rm b}(x,\tau)\to \phi_{\rm FV}
    $ for
  $
    |x|,|\tau|\to \infty .
$
The lifetime of the metastable state is directly controlled by the bounce action $B=S[\phi_{\rm b}]-S[\phi_{\rm FV}]$.

In a local relativistic theory with the analytic $q^2$ term, the spatial and temporal stiffnesses can be made equal by rescaling and the bounce is rotationally symmetric in Euclidean space-time. 
Here, instead, the spatial sector is governed by a fractional nonlocal kernel while imaginary time remains local.
This explicitly breaks the space-time isotropy of the local problem~\cite{defenu2016anisotropic,defenu2017criticality} and leads to anisotropic critical bubbles, as in Fig.~\ref{fig:critical_bubble}. The fractional Laplacian also fixes the asymptotic approach to the false vacuum: writing the displacement from the false vacuum $\varphi=\phi-\phi_{\rm FV}$, the nonlocal saddle equation implies algebraic spatial tails~\cite{classicalInPreparation}, $\varphi(x,\tau)\sim |x|^{-(1+\sigma)}$ as $|x|\to\infty$, in contrast to the exponential tails of a local theory~\cite{coleman1977fate}.

We treat Eq.~\eqref{eq:action_general} as the effective continuum theory governing the universal physics of nonlocal nucleation. Its coefficients can be matched to the static potential and low-momentum spectrum of the microscopic model~\cite{SuppMat}. Here we rescale the field and coordinates to set $c_\tau=c_{\rm LR}=g=\phi_0=1$. Restoring these parameters changes only $h$-independent amplitudes in $B(h)$, and hence leaves all scaling exponents derived below unchanged.

\emph{Anisotropic Derrick identities}---The nonlocal bounce obeys an exact pair of 
scaling identities that generalize Derrick's theorem~\cite{derrick1964comments} to the 
anisotropic Euclidean problem. The bounce action $B$ is a sum of 
three terms, classified by their behavior under coordinate dilations, $B = T_\tau + K_x + U$,  where $T_\tau$ and $K_x$ are equal to the first two terms of~\eqref{eq:action_general}, while $U = \int_{\tau,x}\big[V_{\rm eff}(\phi_{\rm FV} + \varphi)-V_{\rm eff}(\phi_{\rm FV})\big]$. In Coleman's isotropic problem a single dilation suffices to constrain the 
bounce. Here the broken space-time symmetry requires two independent dilations, 
$\varphi(x,\tau)\to\varphi(x/\lambda,\tau/\mu)$, under which 
$T_\tau\to\lambda\mu^{-1}T_\tau$, $K_x\to\lambda^{1-\sigma}\mu\,K_x$, and 
$U\to\lambda\mu\,U$. Stationarity of the action with respect to $\lambda$ and $\mu$ at 
$\lambda=\mu=1$ yields two conditions which fix the energy ratios exactly,
\begin{equation}
\label{eq:virial_ratios}
    \frac{T_\tau}{K_x}=\frac{\sigma}{2},
    \qquad
    \frac{U}{K_x}=\frac{\sigma-2}{2},
\end{equation}
and hence constrain the bounce action to
\begin{equation}
    B = \sigma\,K_x = 2\,T_\tau .
    \label{eq:virial_action}
\end{equation}
Formally, at $\sigma=2$ the spatial kinetic term becomes local and 
Eqs.~\eqref{eq:virial_ratios}--\eqref{eq:virial_action} reduce to the familiar 
two-dimensional Euclidean result $T_\tau=K_x$, $U=0$, $B=2K_x$. Figure~\ref{fig:derrick} shows that these identities are verified numerically (see below) across the full range $0.7 \leq \sigma \leq 2$. These identities 
hold at any tilt $h$ and provide stringent diagnostics of numerical 
convergence.

\begin{figure}
    \centering
    \includegraphics[width=.95\linewidth]{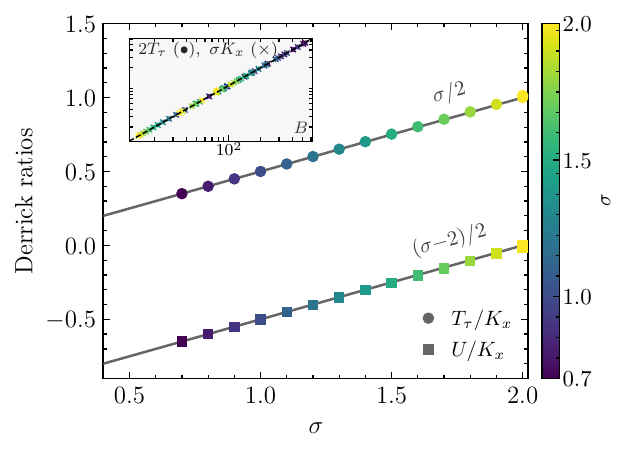}
    \caption{\textbf{Anisotropic Derrick identities.}
    Energy ratios $T_\tau/K_x$ (circles) and $U/K_x$ (squares) 
    extracted from numerical solutions of Eq.~\eqref{eq:bounce_eom} for 
    $\sigma\in[0.7,2]$, colored by $\sigma$. Solid lines are the exact predictions 
    $\sigma/2$ and $(\sigma-2)/2$ of Eq.~\eqref{eq:virial_ratios}. \emph{Inset:} $2T_\tau$ (dots) and $\sigma K_x$ (crosses) versus 
    $B$ on a log-log scale; the collapse onto the diagonal confirms 
    Eq.~\eqref{eq:virial_action} for all $\sigma$ and $h$.}
    \label{fig:derrick}
\end{figure}

\emph{Thin-wall limit}---As in the local theory, we seek an approximate but
analytic description of the bounce in the thin-wall limit, where the energy splitting $\epsilon = V_{\rm eff}(\phi_{\rm FV}) - V_{\rm
eff}(\phi_{\rm TV}) \simeq 2h\phi_0$ is much smaller than the barrier height. In this regime the bounce is a droplet of true vacuum of spatial half-width $R_x$ and imaginary-time half-width $R_\tau$, with narrow wall cores separating the true-vacuum interior from the false-vacuum background,
\begin{equation}
    \varphi(x,\tau) \approx A\,\theta_{\xi_x}(R_x-|x|)\,\theta_{\xi_\tau}(R_\tau-|\tau|),
    \label{eq:droplet_ansatz}
\end{equation}
where $A = \phi_{\rm TV}-\phi_{\rm FV}$ and $\theta_\xi$ is a smoothed step of width $\xi$. The spatial and temporal widths $\xi_x$ and $\xi_\tau$ of the wall cores at $|x|\simeq R_x$ and $|\tau|\simeq R_\tau$ are healing lengths fixed by the curvature of the potential
at the vacua; hence they do not depend on $h$, and only
weakly on $\sigma$.

Evaluating the bounce action $B$ on the ansatz~\eqref{eq:droplet_ansatz} (see End Matter) gives a
thin-wall functional whose extremization fixes $R_x$ through a single
crossover equation,
\begin{equation}
    \epsilon = \beta \,R_x^{-\sigma} + \gamma \,R_x^{-1}. \label{eq:crossover}
\end{equation}
The two terms have distinct physical origins. The term $\beta R_x^{-\sigma}$ comes from the infrared part of the long-range wall energy. In contrast, $\gamma R_x^{-1}$ comes from the wall core: for a sharp interface the fractional energy is UV-divergent for $\sigma > 1$, and the microscopic healing length $\xi_x$ cuts off this divergence, producing an effective surface tension $\gamma$. Moreover, the action scales as $B \propto R_x$, in agreement with Eq.~\eqref{eq:virial_action}, since $T_\tau\propto
R_x$. The entire tilt-dependence of the action is thus inherited from $R_x(h)$,
while the temporal extent satisfies $R_\tau\sim\epsilon^{-1}\sim h^{-1}$
throughout. 

The dominant term in Eq.~\eqref{eq:crossover} changes at $\sigma=1$: for $\sigma<1$ the slower-decaying contribution is $R_x^{-\sigma}$, whereas for $\sigma>1$ it is $R_x^{-1}$. Solving perturbatively,
\begin{equation}
\label{eq:scaling_forms}
    B(h) \simeq
    \begin{cases}
    C_1\,h^{-1/\sigma} + C_2\,h^{-1} & 0<\sigma<1,
    \\[5pt]
    C_1\,h^{-1} (1+c\,\ln h^{-1}) & \sigma=1,
    \\[5pt]
    C_1\,h^{-1} + C_2\,h^{\,\sigma-2} & 1<\sigma<2,
    \end{cases}
\end{equation}
with $C_1>0$ and $C_2<0$. Following the interpretation above, we call the two sides of this crossover the
\textit{IR-dominated} ($0<\sigma<1$) and \textit{UV-dominated} ($1<\sigma<2$) nucleation regimes. The genuinely long-range modification occurs in the IR-dominated regime: the fractional kinetic term directly enters the leading balance, so the wall energy grows with $R_x$ and the leading scaling of the bounce action is $B\sim h^{-1/\sigma}$. On the other hand, the leading energy balance in the UV-dominated regime is controlled by the wall core, as in a local theory, while the infrared part of the long-range wall energy enters the scaling only as a subdominant correction $\sim h^{\sigma-2}$. 
For $\sigma>2$, the analytic $q^2$ term dominates the small-momentum kernel, so the nucleation problem is local from the outset. The asymptotic thin-wall scaling is again the Coleman result $B\sim h^{-1}$, as for exponentially decaying interactions such as those realized in Ref.~\cite{luo2025quantum}. 

Thus, the distinctive metastability induced by long-range interactions sets in only for $\sigma<1$, where the false-vacuum lifetime grows increasingly steeply as $\sigma\to 0$.

\emph{Numerical results}---For local interactions, the bounce solution approaches the false vacuum exponentially fast, thus a thin-wall ansatz appears under control for small $h$. In the long-range problem this control is not automatic. The fractional spatial kernel produces heavy tails, $\varphi(x,\tau)\sim |x|^{-(1+\sigma)}$, so regions far outside the apparent bubble core may still contribute to the action. It is therefore a nontrivial question whether the compact-droplet scaling leading to Eq.~\eqref{eq:scaling_forms} survives for the full saddle.

We answer this question by solving Eq.~\eqref{eq:bounce_eom} directly. Two features make this challenging: the 
fractional Laplacian is nonlocal, and the equation stiffens as $h\to0$, 
where the radii $R_x$ and $R_\tau$ grow while the widths $\xi_x$ and $\xi_\tau$
stay fixed. We discretize space on a Gaussian basis---where the fractional Laplacian is known analytically---and apply Chebyshev collocation in imaginary time. The resulting nonlinear system is solved via Newton-Krylov iteration, using parameter continuation in $h$ to access the stiff small-$h$ regime. Convergence is confirmed against the exact Derrick identities~\eqref{eq:virial_ratios}. See~\cite{SuppMat} for the details.

To test Eq.~\eqref{eq:scaling_forms}, we fit the numerical actions to 
\begin{equation}
\label{eq:numerical_fit_form}
    B_{\rm fit}(h)
    =
    C_1 h^{-\vartheta}
    +
    C_2 f_\sigma(h) + C_{\rm reg},
\end{equation}
where $\vartheta$ is free, while $f_\sigma(h)=h^{-1}$ for $\sigma<1$ and $f_\sigma(h)=h^{\sigma-2}$ for $\sigma>1$, as predicted by the thin-wall analysis. The marginal case $\sigma=1$ is fitted separately to Eq.~\eqref{eq:scaling_forms}. The subleading term is essential over the accessible range of $h$: a pure power law produces systematic deviations~\cite{SuppMat}. 

\begin{figure}
    \centering
    \includegraphics[width=.95\linewidth]{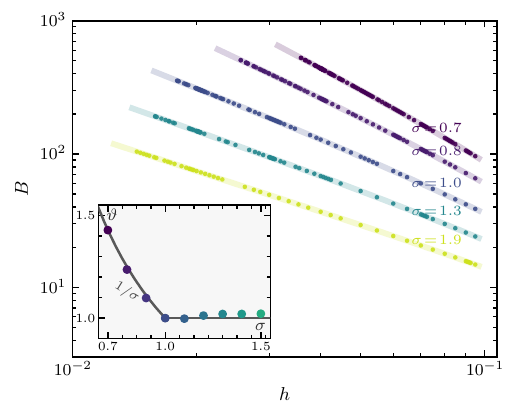}
    \caption{\textbf{Bounce action scaling.}
    Bounce action $B$ versus tilt $h$ for five values of $\sigma$ (colored dots), together with fits of the form~\eqref{eq:numerical_fit_form}
    (shaded bands). The slope steepens as $\sigma$ decreases below one,
    signaling the crossover from UV- to IR-dominated nucleation.
    \emph{Inset:} fitted leading exponent $\vartheta$ versus $\sigma$, in agreement with the two-regime theoretical prediction
    $\vartheta=1/\sigma$ for $\sigma\in(0,1)$ and $\vartheta=1$ for $\sigma\in(1,2)$.}
    \label{fig:scaling}
\end{figure}

As shown in Fig.~\ref{fig:scaling}, Eq.~\eqref{eq:numerical_fit_form} accurately describes the data, and the fitted exponent follows $\vartheta=1/\sigma$ for $\sigma<1$ and $\vartheta=1$ for $\sigma>1$. These results confirm that Eq.~\eqref{eq:crossover} controls
the scaling of the full nonlocal bounce action, not only the simplified
thin-wall ansatz.

\emph{Lattice description}---The field-theory scaling predictions can also be obtained directly from the microscopic Ising chain. In the ordered phase, a bubble of stable vacuum is represented by two kinks separated by $n$ lattice sites. Projecting the spin Hamiltonian~\eqref{eq:Hamiltonian} onto the two-kink sector gives an effective Schrödinger equation for the relative-coordinate wave function $\psi_n$ of a kink pair~\cite{rutkevich1999decay,lagnese2024detecting},
\begin{equation} 
    \sum_{m\geq 1}
    \left(K_{n-m}-K_{n+m}\right)\psi_m +\frac{U(n)}{2}\psi_n =\frac{E}{2}\psi_n , 
    \label{eq:two_kink} 
\end{equation} 
where $K_r=\int_{-\pi}^{\pi}\frac{dq}{2\pi}\, e^{iqr}\varepsilon(q)$ is fixed by the single-kink dispersion $\varepsilon(q)$ in~\eqref{eq:app_kink_dispersion}. The image term enforces $\psi_0=0$. The local transverse field controls kink propagation and gives
\begin{equation} 
    K_r\sim e^{-\kappa_0|r|}|r|^{-3/2}, \qquad \kappa_0=|\ln(|\Gamma|/J_0)|.
    \label{eq:kink_kernel_asymptotic}
\end{equation}
The potential $U(n)=-2|h_z|Mn+U_{\rm LR}(n)$ combines the longitudinal-field gain with the long-range energy of a reversed domain, where $M$ is the spontaneous magnetization and
\begin{equation}
    U_{\rm LR}(n)\simeq \begin{cases} \lambda_\sigma n^{1-\sigma}, & 0<\sigma<1,\\ U_\infty-a_\sigma n^{1-\sigma}, & \sigma>1. \end{cases}
    \label{eq:kink_potential} 
\end{equation} 
Let $n_c$ denote the separation of the nucleating two-kink state resonant with the false vacuum. For $0<\sigma<1$, the growing long-range domain energy balances the bulk gain at $n_c\sim|h_z|^{-1/\sigma}$. For $1<\sigma<2$, the domain energy saturates, so the finite dressed two-kink cost balances the bulk gain at $n_c\sim|h_z|^{-1}$.

Creation from the false vacuum probes microscopic kink separations, whereas the nucleating state is supported near $n_c\gg1$. Within the two-kink approximation, exponentially localized propagation across this separation gives a tunneling suppression $\sim e^{-\kappa_0 n_c}$; see End Matter. Following the approach of Ref.~\cite{rutkevich1999decay}, the decay rate is therefore, to leading exponential accuracy, \begin{equation} B=-\ln\Gamma_{\rm d} \sim 2\kappa_0 n_c. \label{eq:lattice_decay_exponent} \end{equation}
Thus, the lattice problem reproduces the leading regimes of~Eq.\eqref{eq:scaling_forms} without invoking coarse graining.

\emph{Conclusions}---We have established a field-theoretical framework for false-vacuum decay in the presence of long-range interactions, which break the Euclidean spacetime isotropy of Coleman's bounce, inducing algebraic spatial tails and a crossover between two nucleation regimes. The genuinely long-range regime is the IR-dominated one, $0<\sigma<1$, where the infrared part of the wall energy grows with the bubble size and ultimately leads to $B\sim h^{-1/\sigma}$. In this region, long-range interactions strongly suppress nucleation, with the size of the critical bubble scaling steeply as the energy barrier diminishes. The exponent $1/\sigma$ diverges as $\sigma\to0$, beyond which the decay rate is expected to become volume dependent~\cite{ptaszynski2026quantum,xiang2026switching}, vanishing in the thermodynamic limit, as also observed in experiments~\cite{baumgartner2025stability}. For $1<\sigma<2$, the leading balance is UV dominated: the fractional wall energy is cut off by the microscopic wall core and acts as an effective surface tension. The leading Coleman scaling $B\sim h^{-1}$ is therefore recovered, but nonlocality survives in the anisotropic profile, the algebraic tails, and the subleading correction $\sim h^{\sigma-2}$. Numerical solutions of the full saddle confirm this picture and show that subleading terms are essential in the accessible pre-asymptotic regime.

The present work opens the way to the real-time dynamics after nucleation. Once a critical bubble is formed, long-range interactions can affect its growth and interactions with other bubbles~\cite{batini2024real,humar2026resonant}. In sufficiently long-ranged systems (close to $\sigma = 0$) we may even have to go beyond a dilute-bounce picture: in classical long-range models, interactions between droplets have been argued to lead to a condensation of the droplet gas~\cite{mccraw1980metastability}. As $\sigma$ decreases and interactions become longer ranged, two competing phenomena emerge: on the one hand, the single-bubble solution becomes increasingly harder to nucleate; on the other hand, the simultaneous nucleation of multiple (sub-critical) bubbles may trigger an avalanche effect due to bubble-bubble interactions. Whether a bubble-condensed regime exists in quantum long-range false-vacuum decay remains an intriguing open question.

The lattice derivation shows that the continuum scaling laws apply directly to the microscopic Ising chain. Parameter-specific decay rates, beyond these universal exponents, will require a more accurate matching of the continuum potential and stiffnesses to the interacting spin chain, together with tensor-network calculations of its metastable dynamics. We leave this quantitative comparison to future work, particularly for trapped-ion realizations with tunable $\alpha$~\cite{monroe2021programmable}.
Ultimately, our results show that false-vacuum decay provides a sensitive probe of nonlocality in quantum matter. Long-range interactions do not merely renormalize parameters in Coleman's theory: they reshape the critical droplet profile, generating an algebraic spatial structure which, beyond altering the scaling of the decay exponent, should present unmistakable evidence in real-space experiments.

\begin{acknowledgements}
\emph{Acknowledgements:}
We thank M. Delladio, T. Donner, G. Morigi, G. Pagano, G. Perfetto, and A. Solfanelli for fruitful discussions.
 This research was funded by the Swiss National Science Foundation (SNSF) grant numbers 200021--207537, 200021--236722,  and by NCCR SwissMAP. 
\end{acknowledgements}


\appendix
\begingroup
\allowdisplaybreaks

\bibliography{bibquantum}

\clearpage

\section{Thin-wall analysis of the anisotropic bounce}
\label{app:thinwall}

We evaluate the three contributions to the bounce action $T_\tau,K_x,U$ on the
thin-wall droplet ansatz, Eq.~\eqref{eq:droplet_ansatz}. The essential structure
is that the fractional operator acts only in $x$, so at fixed imaginary time the
spatial energy is exactly that of a classical one-dimensional long-range droplet~\cite{classicalInPreparation},
which we may integrate slice by slice over $\tau$.

To calculate the temporal term, we note that the integrand $(\partial_\tau\varphi)^2$ is supported on the two temporal
interfaces at $\tau=\pm R_\tau$. Each is governed by the local operator
$\partial_\tau^2$ and is an ordinary kink of $h$-independent width $\xi_\tau$ and
surface tension $\Sigma_\tau$ (energy per unit length). Each interface spans the
spatial width $2R_x$ of the droplet, so
\begin{equation}
    T_\tau \approx \Sigma_\tau \cdot 2 \cdot (2R_x) = 4\Sigma_\tau R_x.
    \label{eq:app_Ttau}
\end{equation}
At fixed $|\tau|<R_\tau$ the spatial profile is a one-dimensional droplet of
half-width $R_x$. The fractional-Laplacian energy of such a profile is~\cite{classicalInPreparation}
\begin{equation}
    E_{\rm LR}(R_x) \approx 4 \, A^2\!\left\{
    \Sigma_\sigma^{\rm eff} + [\sigma(1-\sigma)]^{-1}  R_x^{1-\sigma}\right\},
    \label{eq:app_E1D}
\end{equation}
where $\Sigma_\sigma^{\rm eff}=[\sigma(\sigma-1)]^{-1}\,\xi_x^{1-\sigma}$ is a wall-core contribution set by
the spatial wall width $\xi_x$---acting as an UV cutoff for $\sigma>1$---and the second term is the IR contribution
growing with the droplet size. Integrating over the duration $2R_\tau$, the spatial term contributes
\begin{equation}
    K_x \approx 4 R_\tau\big(\gamma + \beta\,R_x^{1-\sigma}\big).
    \label{eq:app_Kx}
\end{equation}
The effective tension
$\gamma\equiv 2c_{\rm LR}A^2\Sigma_\sigma^{\rm eff}$ is positive for $\sigma>1$ and negative for $\sigma<1$, since $\Sigma_\sigma^{\rm eff}\propto(\sigma-1)^{-1}$, while $\beta \equiv 2 c_{\rm LR} [\sigma(1-\sigma)]^{-1} A^2$ has
the opposite sign.

Finally, inside the droplet $V_{\rm eff}(\phi)-V_{\rm eff}(\phi_{\rm FV})=-\epsilon$, and
the spacetime volume is $(2R_x)(2R_\tau)$, so
\begin{equation}
    U \approx -\epsilon(2R_x)(2R_\tau) = -4\epsilon \, R_x R_\tau.
    \label{eq:app_U}
\end{equation}
Collecting the three pieces,
\begin{equation}
    B_{\rm tw}(R_x,R_\tau)
    = 4\Sigma_\tau R_x + 4 R_\tau\big(\gamma + \beta R_x^{1-\sigma} - \epsilon R_x\big).
    \label{eq:app_functional}
\end{equation}
Stationarity in $R_\tau$ gives the crossover equation,
\begin{equation}
    \gamma + \beta R_x^{1-\sigma} = \epsilon \, R_x
    \quad\Longleftrightarrow\quad
    \epsilon = \beta \, R_x^{-\sigma} + \gamma \, R_x^{-1},
    \label{eq:app_crossover}
\end{equation}
while stationarity in $R_x$ fixes the temporal extent,
\begin{equation}
    R_\tau = \frac{\Sigma_\tau}{\epsilon-\beta(1-\sigma)R_x^{-\sigma}}
    \sim \epsilon^{-1} .
    \label{eq:app_Rtau}
\end{equation}
Substituting Eq.~\eqref{eq:app_crossover} back into
Eq.~\eqref{eq:app_functional}, the $R_\tau$-proportional terms cancel identically,
leaving
\begin{equation}
    B^\star = 4\Sigma_\tau R_x .
    \label{eq:app_BstarRx}
\end{equation}
The crude box ansatz gives $B^\star = T_\tau$, off by the factor of two of the
exact Derrick relation $B=2T_\tau$ because the box is not the exact saddle. However, the scaling $B\propto R_x$ is exact and ansatz-independent.

At small $\epsilon$ the critical radius $R_x$ is large, so the dominant balance in Eq.~\eqref{eq:app_crossover} is set by the term that decays slowest with $R_x$. The two powers, $R_x^{-\sigma}$ and $R_x^{-1}$, exchange dominance at $\sigma=1$.

For $0<\sigma<1$ the infrared long-range term dominates, $\epsilon\sim \beta R_x^{-\sigma}$, therefore $R_{x,0} \sim\epsilon^{-1/\sigma}$. Perturbing around the leading solution, $R_x=R_{x,0}+\Delta$ gives $\Delta\sim\epsilon^{-1}$, hence
\begin{equation}
    B(h) = C_1 h^{-1/\sigma} + C_2 h^{-1} + \dots,
    \quad C_1>0,\ C_2<0 .
\end{equation}

For $1<\sigma<2$ the leading balance is $\epsilon\sim \gamma R_x^{-1}$, giving $R_x\sim\epsilon^{-1}$. The infrared long-range term $\beta R_x^{-\sigma}$ now gives the first correction to the radius, which scales as $\Delta\sim\epsilon^{\sigma-2}$. Hence,
\begin{equation}
    B(h) = C_1 h^{-1} + C_2 h^{\sigma-2} + \dots,
    \quad C_1>0,\ C_2<0 .
\end{equation}

At $\sigma=1$ the two terms become degenerate, with $R_x^{1-\sigma}/(1-\sigma)\to\ln(R_x/\xi)$, so the crossover equation reads
$\epsilon R_x\sim \ln(R_x/\xi)$, leading to $R_x\sim\epsilon^{-1}\ln(\epsilon^{-1})$
and the marginal form
\begin{equation}
    B(h) = \frac{C_1}{h}\Big(1+c\ln\tfrac1h\Big)+\dots .
\end{equation}
In addition to these singular thin-wall contributions, a genuine regular constant $C_{\rm reg}$ additionally arises from the analytic
$h$-dependence of the prefactors ($A$, $\Sigma_\tau$, $\epsilon$), and is
degenerate with the scaling power only near $\sigma=2$.

The crossover at $\sigma=1$ is specific to nucleation and differs from the equilibrium boundary $\sigma_*^{\rm eq}=2-\eta$, below which long-range interactions modify the critical scaling of the field theory~\eqref{eq:action_general}, where $\eta$ is the anomalous dimension of the short-range Ising chain~\cite{defenu2017criticality}. Hence, for $1<\sigma<\sigma_*^{\rm eq}$, the thermodynamic scaling exponents are $\sigma$-dependent, yet the decay rate remains dominated by an effectively local, surface-tension-like, contribution.

\section{Details on the two-kink description} \label{app:two_kink} 
We outline the microscopic reduction leading to Eq.~\eqref{eq:two_kink}; further details will be presented in Ref.~\cite{pagniBatiniPerfettoDefenuInPreparation}. We separate the nearest-neighbor transverse-field Ising Hamiltonian $H_0$ from the power-law tail and longitudinal field. After duality and Jordan--Wigner fermionization, $H_0$ is quadratic and is diagonalized by a Bogoliubov rotation ~\cite{rutkevich1999decay},
\begin{equation} 
    H_0 = \sum_q \varepsilon(q) \eta_q^\dagger\eta_q, \; \ \varepsilon(q) = 2J_0\sqrt{1+\gamma^2-2\gamma\cos q}, 
    \label{eq:app_kink_dispersion} 
\end{equation}
where $\gamma=|\Gamma|/J_0<1$. The quasiparticles created by $\eta_q^\dagger$ are the dressed kinks of the short-range reference model. We use them as a basis for the full problem and project the power-law tail and the longitudinal field onto the two-kink sector. This description is most controlled deep in the ordered phase, where additional kink pairs remain gapped, and becomes asymptotically accurate for the large bubbles at small $h_z$, whose two walls are widely separated.

We retain the sector containing two such kinks. Translation invariance allows a two-kink state to be resolved into center-of-mass momentum $P$ and relative momentum $q$. Homogeneous decay from the translationally invariant false vacuum involves only $P=0$, for which 
\begin{equation} 
    |\psi\rangle = \sum_q \widetilde\psi(q)\, \eta_q^\dagger\eta_{-q}^\dagger\ket{\mathrm{FV}} = \sum_{n\geq1}\psi_n\ket{n}. 
    \label{eq:app_two_kink_state} 
\end{equation} 
Here $\ket{n}$ denotes a pair of kinks separated by $n$ lattice sites, while $\ket{\mathrm{FV}}$ is the ground state of $H_0$ that will be identified with the false vacuum upon turning on $h_z \neq 0$. The relative-coordinate wave function is the sine transform $\psi_n = \sqrt{2/L} \sum_{q=0}^{\pi} \sin(qn)\,\widetilde\psi(q)$. Its odd continuation, $\psi_{-n}=-\psi_n$, follows from fermionic antisymmetry and enforces $\psi_0=0$. In relative-coordinate space, multiplication by $\varepsilon(q)$ becomes convolution with the kernel $K_r$ defined in the main text. Using the odd continuation on the full line, $\sum_{m\in\mathbb Z}K_{n-m}\psi_m = \sum_{m\geq1} \left(K_{n-m}-K_{n+m}\right)\psi_m$, which explains the kinetic contribution in Eq.~\eqref{eq:two_kink}. Within this projected approximation, kink propagation is inherited from the short-range reference Hamiltonian and is described by Eq.~\eqref{eq:app_kink_dispersion}. Its nearest complex branch points satisfy $q=\pm i\kappa_0$, with $\kappa_0=|\ln\gamma|$. The square-root character of these singularities gives the exponentially localized kernel in Eq.~\eqref{eq:kink_kernel_asymptotic}.

We now project the interaction tail, $r\geq2$, onto this two-kink basis. Since it is diagonal in the $\sigma^z$ basis, it does not move sharp domain walls, so that for two well-separated dressed kinks its leading contribution is a potential depending only on their separation~\cite{liu2019confined}. 
Its large-$n$ form can therefore be obtained by treating the dressed kinks as sharp domain boundaries. For a reversed domain of length $n$, the contribution of the interaction tail is
\begin{equation} 
    U_{\rm LR}(n) = 4J_0\sum_{r\geq2} \frac{\min(r,n)}{r^{1+\sigma}}.
    \label{eq:app_domain_energy} 
\end{equation} 
At large $n$, this yields Eq.~\eqref{eq:kink_potential}, with $\lambda_\sigma=4J_0/[\sigma(1-\sigma)]$ for $\sigma<1$, while $U_\infty=4J_0[\zeta(\sigma)-1]$ and $a_\sigma=4J_0/[\sigma(\sigma-1)]$ for $\sigma>1$. At the marginal point, $U_{\rm LR}(n)\sim \ln n$.

The longitudinal spin operator is nonlocal in the dual fermions: under Kramers--Wannier duality, $\sigma_j^z$ becomes a string. Its projection onto the two-kink sector was derived in Ref.~\cite{rutkevich1999decay}. For a state containing one reversed domain, the string counts the spins between the two kinks and gives the linear contribution $-2|h_z|Mn$, where $M$ is the spontaneous magnetization of the model at $h_z = 0$.

The separation-dependent potential entering the two-kink Schrödinger equation is therefore $U(n) = U_{\rm LR}(n)-2|h_z|Mn$. Thus, up to short-distance dressing of the kink cores, the projected Hamiltonian is approximated by $P_2HP_2 \simeq P_2H_0P_2 +\sum_{n\geq1} U(n) |n\rangle\langle n|$, where $P_2$ is the projector onto the two-kink subspace, giving Eq.~\eqref{eq:two_kink}. 

It remains to connect the resonant two-kink state to false-vacuum decay. Let $|\psi^{(\ell)}\rangle$ denote a two-kink eigenstate of the projected Hamiltonian, where $\ell$ labels the resonant levels.
The longitudinal-field string has both diagonal and kink-number-changing matrix elements: its diagonal projection gives the linear potential, while its leading off-diagonal component couples the false vacuum to the two-kink sector~\cite{rutkevich1999decay}. 
With $H_{h_z}=-h_z\sum_i\sigma_i^z$, the corresponding matrix element is \begin{equation} \mathcal M_\ell = \langle\psi^{(\ell)}|H_{h_z}|\mathrm{FV}\rangle = \sum_{n\geq1} V_n\psi_n^{(\ell)*}, \quad V_n = \langle n|H_{h_z}|\mathrm{FV}\rangle . \label{eq:app_decay_matrix_element} \end{equation}
The pair-creation vertex is exponentially localized at microscopic kink separations. Deep in the ordered phase~\cite{rutkevich1999decay,pagniBatiniPerfettoDefenuInPreparation} \begin{equation} V_n \propto \gamma^n = e^{-\kappa_0n}, \qquad \kappa_0=|\ln\gamma|, \label{eq:app_pair_vertex} \end{equation} up to subexponential and $n$-independent factors. Consequently, for a resonant state concentrated near $n_c\gg1$, Eq.~\eqref{eq:app_decay_matrix_element} probes its wave-function tail at microscopic separations $n=O(1)$.

Propagation from the nucleating bubble to this microscopic region is governed by the exponentially localized kinetic kernel \eqref{eq:kink_kernel_asymptotic}. The resulting wave-function envelope is $-\ln | \psi_{n=O(1)}^{(\ell)} | \sim \kappa_0n_c$, and therefore $-\ln|\mathcal M_\ell| \sim  \kappa_0n_c$. Physically, each propagation path connecting the two regions spans a total distance of at least $n_c$, while paths involving backtracking are more strongly suppressed. A detailed derivation will be presented in Ref.~\cite{pagniBatiniPerfettoDefenuInPreparation}. Since $\Gamma_{\rm d}\propto|\mathcal M_\ell|^2$, these relations reproduce the two scaling laws stated in the main text.

\onecolumngrid

\appendix
\clearpage
\section*{Supplemental Material}

\setcounter{equation}{0}
\renewcommand{\theequation}{S\arabic{equation}}

\setcounter{figure}{0}
\renewcommand{\thefigure}{S\arabic{figure}}

This Supplemental Material provides: (i) two complementary derivations of the nonlocal continuum theory from the microscopic Ising chain, together with an approximate matching of its coefficients; (ii) the numerical method used to solve the nonlocal saddle-point equation; (iii) validation against the anisotropic Derrick identities; (iv) a comparison between pure power-law and regime-dependent scaling fits; and (v) additional bounce profiles across the IR--UV crossover.

\section{Mapping the long-range Ising chain to a field theory}
\label{app:mapping}

Here we outline the mapping from the long-range mixed-field Ising chain to the spatially nonlocal Euclidean field theory used in the main text; see also~\cite{suzuki1976generalized, hertz1976quantum,sachdev2011quantum,defenu2017criticality}.

\paragraph{Suzuki--Trotter decomposition.}
We start from the Hamiltonian
\begin{equation}
\label{eq:app_LR_TFIM}
    H =
    -\sum_{i<j} V_{ij}\sigma_i^z\sigma_j^z
    -h_z\sum_i\sigma_i^z -\Gamma\sum_i\sigma_i^x,
\end{equation}
which retrieves main-text Eq.~(1) for $V_{ij}=J_0 r_{ij}^{-\alpha}$. We split $H=H^{(z)}+H^{(x)}$, where $H^{(z)}$ represents the two longitudinal terms in~\eqref{eq:app_LR_TFIM}, while $H^{(x)}$ is the transverse-field term. The partition function at temperature $k_B T = \beta^{-1}$ is
\begin{equation}
    Z
    =
    \lim_{M\to\infty}
    \Tr
    \left[
    e^{-\Delta\tau H^{(z)}}
    e^{-\Delta\tau H^{(x)}}
    \right]^M,
    \quad
    \Delta\tau=\frac{\beta}{M} .
\end{equation}
Inserting complete sets of $\sigma^z$ eigenstates at each time slice gives a
classical statistical model for Ising variables $s_{i,n}=\pm1$, where $n$
labels imaginary-time slices. The longitudinal part is diagonal, whereas the transverse-field matrix element can be written in the standard nearest-neighbor
form along imaginary time,
\begin{equation}
    \mel{s_{i,n}}{e^{\Delta\tau\Gamma\sigma_i^x}}{s_{i,n+1}}
    =
    e^{K_\tau s_{i,n}s_{i,n+1}+C_\tau},
\end{equation}
with $K_\tau = -\frac{1}{2}\log\tanh(\Gamma\Delta\tau)$ 
and $C_\tau$ an irrelevant constant. Thus, we obtain a classical Ising model in $d+1$ Euclidean dimensions, described by
\begin{equation}
    Z
    \propto
    \sum_{\{s_{i,n}\}}
    e^{-S_T[s]},
\end{equation}
i.e.~by the action $S_T[s]
    =
    -\sum_n
    (
        \Delta\tau\sum_{i<j}V_{ij}s_{i,n}s_{j,n}
        +
        K_\tau\sum_i s_{i,n}s_{i,n+1}
        +
        \Delta\tau h_z\sum_i s_{i,n})$,
with long-range spatial couplings.

\paragraph{Hubbard--Stratonovich representation.}
It is useful to combine the spatial site $i$ and time slice $n$ into a single
index $y=(i,n)$. The quadratic ferromagnetic part of $S_T$ can be written schematically as $-\frac{1}{2}\sum_{y,y'} s_y M_{y,y'}s_{y'}$, where $M$ contains both the spatial interaction and the nearest-neighbor
imaginary-time coupling. Introducing a real Hubbard--Stratonovich field
$\phi$ gives
\begin{equation}
    e^{\frac{1}{2}s M s}
    \propto
    \int \mathcal D\phi\,
    \exp\left[
        -\frac{1}{2}\phi M^{-1}\phi
        +
        \phi s
    \right],
\end{equation}
where matrix notation is used. After summing independently over the Ising
variables, one obtains
the Hubbard-Stratonovich effective action 
\begin{equation}
\label{eq:app_HS_action_explicit}
\begin{aligned}
    S_{\rm HS}[\phi]
    =
    &\frac{1}{2}
    \sum_{y,y'}
    \phi_y(M^{-1})_{y,y'}\phi_{y'}
    \\
    & -
    \sum_y
    \log\left[
        2\cosh(\phi_y+\Delta\tau h_z)
    \right].
\end{aligned}
\end{equation}
The continuum order-parameter field used in the main text is the coarse-grained
version of this Hubbard--Stratonovich field.

\paragraph{Continuum limit and fractional spatial kernel.}

To obtain the long-wavelength field theory, we diagonalize the quadratic kernel
in momentum and frequency space (assuming $\beta \to \infty$ hereafter). 
At fixed nonzero Trotter spacing $\Delta\tau$, the quadratic interaction matrix is diagonal in momentum and Euclidean frequency and has the form
\begin{equation}
    M(q,\omega_E)
    =
    \Delta\tau V(q)
    +
    2K_\tau\cos(\omega_E\Delta\tau)
    +
    \lambda ,
\end{equation}
where $\lambda$ denotes a possible diagonal shift chosen so that the Gaussian
kernel is positive. At fixed $\Delta\tau$, expanding the temporal dependence at small frequency, $\cos(\omega_E\Delta\tau) =1-\omega_E^2\Delta\tau^2/2+\dots$, produces the analytic $\omega_E^2$ structure of a local temporal kinetic term. In contrast, the spatial part is controlled by the Fourier transform of the power-law
interaction. For $V(r)\sim \frac{1}{r^{d+\sigma}}$, with $\sigma=\alpha-d$,  the small-momentum expansion contains both analytic and nonanalytic terms,
\begin{equation}
\label{eq:app_Vq_expansion}
    V(q)
    =
    V(0)
    -
    v_\sigma |q|^\sigma
    -
    v_2 q^2
    +
    \dots .
\end{equation}
The nonanalytic contribution is the characteristic signature of the long-range
interaction. Provided that the shifted uniform kernel is invertible, its formal long-wavelength expansion at fixed $\Delta\tau$ has the form
\begin{equation}
\label{eq:app_inverse_kernel}
    M^{-1}(q,\omega_E)
    \simeq
    r_0(\Delta\tau) + a_\tau(\Delta\tau)\omega_E^2 + a_\sigma(\Delta\tau)|q|^\sigma + a_2(\Delta\tau)q^2 + \dots .
\end{equation}
The coefficients at this stage depend on $\Delta\tau$, on the diagonal shift $\lambda$, and on the normalization of the Hubbard--Stratonovich field. Transforming back to
real space and imaginary time yields the continuum action
\begin{equation}
\label{eq:app_continuum_action}
\begin{aligned}
    S_{\rm eff}[\phi]
    =
    &\int_{\tau, x}
    \bigg[
    \frac{c_\tau}{2}(\partial_\tau\phi)^2
    +
    \frac{c_{\rm LR}}{2}
    \phi(-\partial_x^2)^{\sigma/2}\phi
    +
    \frac{c_{\rm SR}}{2}(\partial_x\phi)^2
    + 
    V_{\rm eff}(\phi)
    \bigg].
\end{aligned}
\end{equation}
Here $(-\partial_x^2)^{\sigma/2}$ is defined by its Fourier representation,
$|q|^\sigma$. The coefficients $c_\tau$, $c_{\rm SR}$, and $c_{\rm LR}$ depend
on the microscopic parameters and on the normalization of $\phi$.

A caveat concerns the normalization of these coefficients. The continuous-time limit is strongly anisotropic: since \begin{equation} K_\tau = -\frac{1}{2}\log\tanh(\Gamma\Delta\tau) \simeq -\frac{1}{2}\log(\Gamma\Delta\tau), \end{equation} the temporal coupling diverges as $\Delta\tau\to0$. This divergence is not pathological, but enforces continuity between neighboring imaginary-time slices as their separation vanishes; see also~\cite{kogut1979introduction}. Consequently, the finite-$\Delta\tau$ coefficients in Eq.~\eqref{eq:app_inverse_kernel} cannot be identified directly with those of Eq.~\eqref{eq:app_continuum_action} by simply sending $\Delta\tau\to0$. Their quantitative determination requires a controlled anisotropic continuum limit and an appropriate normalization of the coarse-grained field. This subtlety in extracting continuum nucleation quantities from the Trotter representation has also been emphasized in Ref.~\cite{moss2025notes}. It affects the nonuniversal coefficients, but not the operator content identified above: the temporal sector is analytic and begins with $\omega_E^2$, while the power-law interaction produces the nonanalytic spatial term $|q|^\sigma$. Below, we match the continuum coefficients independently from the quantum Hamiltonian within linear spin-wave theory. For the short-range Ising chain, the corresponding continuous-time connection can be controlled using the exact solution of the anisotropic two-dimensional classical Ising model~\cite{rutkevich1999decay,moss2025notes}; no analogous exact construction is available for the long-range model considered here.

For $0<\sigma<2$, the nonanalytic $|q|^\sigma$ term dominates the analytic
$q^2$ stiffness at small momentum. This is the fractional-dominated regime
considered in the main text. There we set $c_{\rm SR}=0$ and use the pure
fractional representative of the long-wavelength universality class. In one
spatial dimension this corresponds to $1<\alpha<3$. For $\sigma>2$, the analytic $q^2$ stiffness dominates the infrared spatial
dynamics and the leading continuum theory becomes local. 

\paragraph{Effective potential.}
The local part of the Hubbard--Stratonovich action is generated by the spin
trace. For a uniform field, it has the schematic form
\begin{equation}
    V_{\rm loc}(\phi)
    =
    -\log\left[
        2\cosh(\phi+\Delta\tau h_z)
    \right]
    +
    \frac{\phi^2}{2M_0},
\end{equation}
up to additive constants and field-normalization factors, where $M_0$ denotes
the uniform part of the quadratic kernel. Near the symmetry-breaking transition,
this local potential can be expanded in powers of $\phi$ and $h_z$. After coarse
graining and field rescaling, the minimal form needed to describe false-vacuum
decay is the tilted double well
\begin{equation}
\label{eq:app_tilted_quartic}
    V_{\rm eff}(\phi)
    =
    \frac{g}{4}
    (\phi^2-\phi_0^2)^2
    -
    h\phi .
\end{equation}
Here $h$ is the continuum tilt induced by the microscopic longitudinal field
$h_z$, up to nonuniversal normalization factors. The precise polynomial form of
$V_{\rm eff}$ is not essential for the scaling theory. What is required is the
presence of two locally stable minima separated by a barrier and a small
vacuum-energy splitting
\begin{equation}
    \epsilon
    =
    V_{\rm eff}(\phi_{\rm FV})
    -
    V_{\rm eff}(\phi_{\rm TV})
    >0 .
\end{equation}
This justifies the phenomenological action used in the main text, even far from the symmetry-breaking transition.

\section{Mean-field theory and quadratic bosonic approximation}
In this section we develop the mean-field treatment of the Hamiltonian main-text Eq.~(1), and a subsequent quadratic bosonic theory which is obtained by linearizing the Holstein-Primakoff transformation. This provides an alternative route to obtaining the Euclidean action main-text Eq.~(2). 
\subsection{Mean-field theory}
We assume a product state of spin-coherent states at each site with a common Bloch vector $\boldsymbol{m} = (m_x,m_y,m_z) = (\sin \theta, 0, \cos\theta)$, parametrized by the angle $\theta\in[0,\pi]$. This homogeneous ansatz allows the decoupling of the Ising interaction, so that the energy per site obtained from main-text Eq.~(1) reads
\begin{equation}
    V_{\rm MF}(\phi) = -\frac{\tilde J(q=0)}{2} m_z^2 - \Gamma m_x - h_z m_z = -\frac{\tilde J(0)}{2} \phi^2 - \Gamma \sqrt{1-\phi^2} - h_z \phi,
    \label{eq:MF_potential}
\end{equation}
where $\phi \coloneqq m_z \in [-1,1]$. Here, we have introduced
\begin{equation}
   \tilde J(q)=\sum_{r\neq0}\frac{J_0}{|r|^\alpha}e^{iqr}=2J_0\sum_{r\ge1}\frac{\cos qr}{r^\alpha} = \tilde J(0)-v_\sigma|q|^{\sigma}-v_2q^2+\dots, \qquad \tilde J(0) = 2J_0 \zeta(\alpha),
   \label{eq:Fourier_transform_LR_interaction}
\end{equation}
where $v_\sigma>0, v_2=J_0\zeta(\sigma-1)$ in the range $\sigma = \alpha-1 \in(0,2)$.
When $h_z = 0$, the possible minima of the potential are at $\phi=0$ (paramagnet) or $\phi=\pm\phi_\star$ (ferromagnet), with $\phi_\star = \sqrt{1-(\Gamma/\tilde J(0))^2}$, thus requiring the transverse field to lie in the ordered phase $|\Gamma|< \tilde J(0) \eqqcolon \Gamma_c$. When $h_z>0$ is small, we denote the new minima by $\phi_{\rm FV} = -\phi_\star + O(h_z)$ and $\phi_{\rm TV} = \phi_\star + O(h_z)$, with a slight abuse of notation. Their splitting is
\begin{equation}
    \epsilon_{\rm MF} \coloneqq V_{\rm MF}(\phi_{\rm FV})-V_{\rm MF}(\phi_{\rm TV}) = 2h_z \phi_\star + O(h_z^3).
\end{equation}
Expanding the term $\sqrt{1-\phi^2}$ in \eqref{eq:MF_potential}, we obtain
\begin{equation}
    V_{\rm MF}(\phi) = - \Gamma +\frac{\Gamma-\tilde J(0)}{2}\phi^2+\frac{\Gamma}{8} \phi^4 - h_z \phi +O(\phi^6),
\end{equation}
which, up to additive constants, can be matched to the field theory potential $V_{\rm eff}(\phi) = \tfrac{g}{4}(\phi^2-\phi_0^2)^{2} - h \phi$ used in the main text, by means of the identifications
\begin{equation}
    h = h_z, \qquad g = \Gamma/2, \qquad \phi_0^2 = 2(\tilde J(0)/\Gamma-1).
\end{equation}
Note that the precise form of the mean-field potential is different from the one obtained via the Hubbard-Stratonovich approach, but the shape of a double-well potential with a barrier $\Delta V = V(0)-V(\phi_{\rm FV}) > 0$ is independent of the route we choose to write a continuous theory. In order to connect the field theory results more closely to the microscopic Ising chain, we may choose its potential to be $V_{\rm eff}(\phi) \equiv V_{\rm MF}(\phi)$, which can be implemented with minimal changes within our numerical method to solve the nonlocal saddle point equation.

Note that the sign of the transverse field is immaterial. The global unitary transformation $U=\prod_i\sigma_i^z$ leaves the Ising interaction and the longitudinal-field term invariant, while $U\sigma_i^xU^\dagger=-\sigma_i^x$. Hence, \begin{equation} U H(\Gamma)U^\dagger=H(-\Gamma), \end{equation} so the two signs of $\Gamma$ are unitarily equivalent. We may therefore take $\Gamma>0$ without loss of generality.
\subsection{Quadratic bosonic approximation}
In order to extract a microscopic expression for the kinetic coefficients of the field theory main-text Eq.~(2) as well, we allow the spins to fluctuate around the ordered state. In particular, we are interested in the frequency dependence at small momentum $q$ of such fluctuations, or magnons. In the following, we set $h_z=0$ when extracting the kinetic coefficients, since a small $h_z$ slightly shifts the mean-field saddle but does not alter the nonanalytic small-$q$ structure relevant here.

We consider a change of basis in which the new quantization axis points along the Bloch vector $\boldsymbol{m}$, so that $\langle \tilde \sigma^z \rangle = 1$. This is achieved by a rotation about the original $y$-axis by the angle $\theta$ satisfying $\cos\theta = \phi_\star$ and $\sin\theta =\sqrt{1-\phi_\star^2} = |\Gamma|/\tilde J(0)$:
\begin{equation}
    \sigma_i^z = \cos\theta \, \tilde\sigma_i^z - \sin\theta \, \tilde\sigma_i^x,
    \qquad 
    \sigma_i^x = \sin\theta \, \tilde\sigma_i^z + \cos\theta \, \tilde\sigma_i^x.
\end{equation}
At this stage, we may introduce the bosonic operators $a_i,a_i^\dagger$, and the Holstein-Primakoff mapping for spin-$1/2$ systems:
\begin{equation}
    \tilde\sigma_i^z = 1-2n_i, \qquad \tilde\sigma^+_i=\tilde\sigma_i^x+i\tilde\sigma^y_i = 2\sqrt{1-n_i}\;a_i,\qquad \tilde\sigma^-_i = \tilde\sigma_i^x-i\tilde\sigma^y_i = 2\,a_i^\dagger\sqrt{1-n_i},
\end{equation}
where $n_i = a_i^\dagger a_i$. Our approximation, corresponding to linear spin-wave theory, is to neglect all terms that are more than quadratic in the bosonic operators. In particular, $\tilde\sigma_i^x \simeq a_i + a_i^\dagger$ and $\tilde\sigma_i^z = 1-2 a_i^\dagger a_i$. Inserting these into the long-range Ising Hamiltonian main-text Eq.~(1) yields a zeroth order term equal to the classical mean-field energy, $E_0/N = V_{\rm MF}(\theta)$, a vanishing first-order term -- due to the stationary value $\sin\theta = |\Gamma|/\tilde J(0)$ -- and a quadratic Hamiltonian of the form
\begin{equation}
    H_2=2\tilde J(0)\sum_i a_i^\dagger a_i - \frac{1}{2}\sin^2\theta\sum_{i\neq j}J_{ij}\, (a_i+a_i^\dagger)(a_j+a_j^\dagger).
\end{equation}
Equivalently, going to momentum space via $a_i = N^{-1/2} \sum_q e^{iqr_i} a_q$,
\begin{equation}
    H_2=\sum_q\Big[A_q\,a_q^\dagger a_q +\frac{B_q}{2}\big(a_qa_{-q}+a_q^\dagger a_{-q}^\dagger\big)\Big]+\text{const}, \qquad 
    A_q=2\tilde J(0)-\sin^2\theta\,\tilde J(q),\quad B_q=-\sin^2\theta\,\tilde J(q),
\end{equation}
making apparent that this approximation, unlike mean-field theory, retains information about the spatial structure of the excitations.

As per the standard procedure, to eliminate the anomalous terms proportional to $B_q$, we perform a Bogoliubov rotation
\begin{equation}
    a_q=u_q b_q+v_q b_{-q}^\dagger,\qquad u_q=u_{-q},\ v_q=v_{-q},\qquad u_q^2-v_q^2=1,
\end{equation}
from the $a$-bosons to $b$-bosons. The calculation is straightforward and yields $H_2 = \sum_q \omega(q) \, b_q^\dagger b_q + \mathrm{const}$, where the dispersion relation is
\begin{equation}
    \omega(q) = \sqrt{A_q^2-B_q^2} = 2 \sqrt{\tilde J(0) [\tilde J(0)-\sin^2\theta \, \tilde J(q)]},
\end{equation}
with the correct limits of a flat band for $\Gamma \to 0$ (i.e.  $\sin \theta \to 0$) and a vanishing gap $\Delta \coloneqq \omega(q=0) = 2\tilde J(0)\cos\theta$ as we approach the critical point $\Gamma \to \Gamma_c = \tilde J(0)$ (i.e. $\cos \theta \to 0$). Using the small-$q$ expansion of $\tilde J(q)$ in Eq. \eqref{eq:Fourier_transform_LR_interaction}, we find
\begin{equation}
    \omega(q)^2 = \Delta^2 + \frac{4 \Gamma^2}{\tilde J(0)} (v_\sigma|q|^{\sigma}+v_2q^2+\dots).
\end{equation}
\subsection{Matching between field theory and spin chain}
We can now match the coefficients of the continuum theory directly from the spin-wave dispersion. Expanding the field around either minimum, $\phi=\phi_{\rm vac}+\delta\phi$, the quadratic part of the action main-text Eq.~(2), including a local spatial term, reads 
\begin{equation} 
    S_2 = \frac{1}{2} \int_{\omega_E,q} \left[ c_\tau\omega_E^2 + V_{\rm eff}''(\phi_{\rm vac}) + c_{\rm LR}|q|^\sigma + c_{\rm SR}q^2 +\dots \right] |\delta\phi(\omega_E,q)|^2 . 
\end{equation}
After analytic continuation to real frequency, $\omega_E\to-i(\omega+i0^+)$, the excitation spectrum is determined by the zeros of the inverse retarded propagator. Therefore,
\begin{equation} 
    \omega(q)^2 = \frac{V_{\rm eff}''(\phi_{\rm vac})}{c_\tau} + \frac{c_{\rm LR}}{c_\tau}|q|^\sigma + \frac{c_{\rm SR}}{c_\tau}q^2 +\dots .
\end{equation}
Comparing this expression with the spin-wave result obtained above gives 
\begin{equation} 
    \frac{V_{\rm eff}''(\phi_{\rm vac})}{c_\tau} = \Delta^2, \qquad \frac{c_{\rm LR}}{c_\tau} = \frac{4\Gamma^2}{\tilde J(0)}v_\sigma, \qquad \frac{c_{\rm SR}}{c_\tau} = \frac{4\Gamma^2}{\tilde J(0)}v_2. 
    \label{eq:coefficient_ratios} 
\end{equation}
The dispersion fixes the ratios in Eq.~\eqref{eq:coefficient_ratios} independently of the field normalization. If the continuum field is identified with the mean-field magnetization and the full local potential is chosen as $V_{\rm eff}=V_{\rm MF}$, the remaining normalization is fixed by $V_{\rm eff}''(\phi_{\rm vac}) = V_{\rm MF}''(\phi_\star) = \Delta^2 \, \tilde J(0)/(4\Gamma^2)$, giving
\begin{equation}
    c_\tau=\tilde J(0)/(4\Gamma^2), \qquad c_{\rm LR}=v_\sigma, \qquad c_{\rm SR}=v_2.
    \label{eq:SM_absolute_coefficients} 
\end{equation}
Thus, mean-field theory fixes both the local potential and the normalization of the temporal term, while the nonanalytic and analytic parts of the microscopic interaction directly determine the spatial coefficients of the continuum theory in main-text Eq.~(2).

Mean-field and linear spin-wave theory are standard tools for analyzing spin systems in experimental implementations. Here we use them to construct an explicit matching to the nonlocal continuum theory. This matching can be refined systematically by replacing the mean-field potential and linear-spin-wave dispersion with numerically calculated or experimentally measured lattice quantities.

\section*{Numerical methods}
\label{sec:method}

\subsection{Boundary-value problem}
\label{sec:model}

The continuum action and saddle-point equation are defined in the main text.
For completeness, we first state the boundary-value problem solved
numerically. We work with the shifted field
$\varphi=\phi-\phi_{\rm FV}$, so that the false vacuum corresponds to
$\varphi=0$. The bounce equation is
\begin{equation}
  \Big[\,{-c_\tau}\,\partial_\tau^2 + c_{\rm LR}\,\Frac{\sigma/2} \Big]\varphi
  + V'(\varphi+\phi_{\rm FV}) = 0,
  \label{eq:eom}
\end{equation}
with the tilted double well $V(\phi)=\tfrac{g}{4}(\phi^2-\phi_0^2)^2-h\phi$. Unless stated otherwise, all numerical results are obtained in the units
$c_\tau=c_{\rm LR}=g=\phi_0=1$. The parameter scan is therefore over
the bias $h$ and the long-range exponent $\sigma$.

On the infinite Euclidean plane, the bounce is reflection-symmetric about
the droplet center, i.e. even in both $x$ and $\tau$, and approaches the false vacuum as
$|x|,|\tau|\to\infty$. Because the spatial operator is nonlocal, this
approach is algebraic in space,
$\varphi(x,0)\sim |x|^{-(1+\sigma)}$, while it is exponential in imaginary
time. Numerically, we solve the problem on the finite box
$(x,\tau)\in[-L_x,L_x]\times[-L_\tau,L_\tau]$, choosing $L_x$ and
$L_\tau$ large enough that the finite-domain truncation does not affect the
bounce action within our convergence tolerance. The reflection symmetries
are built directly into the spatial and temporal discretizations described
below, allowing us to work only on the quadrant $x,\tau\ge 0$.

 At the outer temporal boundary we impose the
Dirichlet condition
$
\varphi(x,L_\tau)=0 ,
$
which becomes essentially exact once the exponential temporal decay is well resolved before
reaching $L_\tau$. The only condition that cannot be imposed exactly on a
finite domain is the algebraic spatial tail. We therefore truncate it at
$x=\pm L_x$, and check convergence in $L_x$ to ensure that this
truncation does not bias the action. Since the bounce is a saddle point of the Euclidean action, not a minimum,
we solve \eqref{eq:eom} as a nonlinear root-finding problem.

\subsection{Spatial discretization}
 At fixed imaginary time, we discretize the spatial dependence by expanding
the shifted field as a sum of $N_x$ smooth,
localized Gaussian basis functions. The Gaussian centers lie on the cell-centered grid
\begin{equation}
    x_j=\left(j-\tfrac12 \right)\Delta x, \quad j=1,\dots,N_x, 
\end{equation} with
$\Delta x=L_x/N_x$. 
To impose the reflection symmetry about the bubble center exactly, each basis
function is symmetrized as an even pair,
\begin{equation}
    G_j(x)
    =
    e^{-w_g^2(x-x_j)^2}
    +
   e^{-w_g^2(x+x_j)^2}
 ,
\end{equation}
where $w_g$ is the inverse width of the Gaussian basis functions. In this way, 
\begin{equation}
    \varphi(x,\tau)
    =
    \sum_{j=1}^{N_x} c_j(\tau)\,G_j(x)
\end{equation}
 enforces the spatial reflection symmetry exactly. Since
each $G_j(x)$ is even in $x$, any linear combination satisfies
$\partial_x\varphi(0,\tau)=0$, independently of the coefficients
$c_j(\tau)$.
In the discretized problem, the coefficients are stored in a matrix
$C_{j\ell}$, where $\ell$ labels the imaginary-time node. The field values on the spatial grid are obtained by applying the interpolation matrix
$\Bop$,
\begin{equation}
    \varphi_{i\ell}=(\Bop C)_{i\ell} =\sum_{j=1}^{N_x} G_j(x_i)\,C_{j\ell}.
\end{equation}
Because $G_j(x_i)$ contains both $x_i-x_j$ and the mirror distance
$x_i+x_j$, the operator $\Bop$ has a Toeplitz-plus-Hankel structure. This
structure allows $\Bop$ to be applied efficiently by FFTs, without forming
the full $N_x\times N_x$ matrix explicitly.

The Gaussian basis is also useful for applying the fractional Laplacian.
Since the operator is linear, acting on the expanded field gives
\begin{equation}
  \Frac{\sigma/2}\varphi(x,\tau)
  =
  \sum_{j=1}^{N_x} c_j(\tau)\,
  \Frac{\sigma/2}G_j(x).
\end{equation}
Thus the problem reduces to applying the fractional Laplacian to a single
Gaussian. This can be done analytically:
\begin{equation}
  \Frac{\sigma/2}e^{-w_g^2 r^2}
  =
  \frac{
  w_g^{\sigma}4^{\sigma/2}
  \Gamma\!\left(\frac{\sigma+1}{2}\right)
  }{\sqrt{\pi}}\,
  {}_1F_1\!\left(
  \frac{\sigma+1}{2};
  \frac12;
  -w_g^2 r^2
  \right),
  \label{eq:kernel}
\end{equation}
which follows from the spectral definition of the fractional Laplacian~\cite{kwasnicki2017ten,lischke2020fractional},
\begin{equation}
    \Frac{\sigma/2}f(x) = \int\frac{\mathrm{d}k}{2\pi}\,e^{ikx}|k|^\sigma\tilde f(k).
\end{equation}
Applied to $G_j$, $r$ is either $x-x_j$ or $x+x_j$, so evaluating this at the grid points gives exactly the Toeplitz-plus-Hankel kernel found for the interpolation operator $\Bop$. We apply this operator by
FFT-based convolutions, without forming the full $N_x\times N_x$ matrix.
All imaginary-time slices are processed together as columns of the coefficient
matrix $C_{j\ell}$.

The remaining numerical choice is the Gaussian width parameter. We write it
as the dimensionless ratio $q=w_g\Delta x$. This choice requires some care---a balance familiar from radial-basis-function methods more generally~\cite{fornberg2011stable}. 
Small $q$ gives broad, strongly overlapping Gaussians, which improves
smoothness but can make the basis ill-conditioned. Large $q$ gives narrow
Gaussians, which improves conditioning but reduces the overlap between
neighboring grid points. We use $q=0.60$, which gives a stable compromise for
all of our calculations.

\subsection{Temporal discretization}
In imaginary time we use a spectral collocation grid based on
Chebyshev--Lobatto points. These points cluster near the interval endpoints
and efficiently resolve smooth, rapidly varying profiles
with relatively few nodes, and far more efficiently than a uniform grid~\cite{boyd2001chebyshev,trefethen2000spectral}. On the full interval $[-L_\tau,L_\tau]$, the nodes are
\begin{equation}
\tau_k=L_\tau\cos\!\left(\frac{\pi k}{N_\tau}\right),
\qquad k=0,\ldots,N_\tau ,
\end{equation}
with $N_\tau$ even. The second-derivative matrix $D^{(2)}$ is obtained
from the standard Chebyshev differentiation matrix. We rescale it to the
interval $[-L_\tau,L_\tau]$.

Since the bounce is even in $\tau$, the values on the negative half of the
grid are not independent: $\varphi_{N_\tau-k}=\varphi_k$.  We therefore fold the full Chebyshev grid
onto the half-domain $\tau\ge0$. We keep only the $M_t=N_\tau/2+1$ nodes $k=0,\dots,N_\tau/2$. 
With our ordering, $k=0$ is the outer wall $\tau=L_\tau$. The last node,
$k=M_t-1$, is the center $\tau=0$. The folded second-derivative operator is
\begin{equation}
  D_\tau^{(2)}{}_{ik}=
  \begin{cases}
    D^{(2)}_{ik}+D^{(2)}_{i,N_\tau-k}, & k=0,\dots,M_t-2,\\
    D^{(2)}_{i,M_t-1}, & k=M_t-1.
  \end{cases}
  \label{eq:foldedD}
\end{equation} 
Equivalently, each column is added to its mirror partner, except for the
center column, which has none. Applying $D_\tau^{(2)}$ to the half-grid
values reproduces the action of $D^{(2)}$ on the evenly extended full grid. The Neumann condition $\partial_\tau\varphi(x,0)=0$ is therefore built into
the discretization, rather than imposed as an additional equation.

Clenshaw--Curtis quadrature weights are folded in the same way and used for
temporal integration. 
Unlike the spatial operators, $D_\tau^{(2)}$ is a dense
$M_t\times M_t$ matrix. This remains inexpensive in practice, since
$M_t\leq 65$ is in general  sufficient to resolve the bounce in our calculations.

\subsection{Nonlinear solver}

As previously noted, the bounce is a saddle point of the action, so we cannot
minimize the latter. Instead, we write the discretized version of
Eq.~\eqref{eq:eom} as a residual $\vF(C)=0$, where $C$ collects the field's
expansion coefficients, and solve for the $C$ that makes this residual
vanish using Newton's method~\cite{knoll2004jacobianfree}. Newton's method
converges quickly once close to the solution but can fail if started too far
from it, so we first take a number of gradient-descent-like steps
(L-BFGS~\cite{liu1989limited}) to approach the solution, before switching to
Newton's method to converge to high precision. Each Newton step requires
solving a large linear system, which we do approximately using the GMRES
algorithm~\cite{saad1986gmres,eisenstat1996choosing}. To accelerate this
step we use a preconditioner built from the exact momentum-space symbol
$k^\sigma$ of the fractional Laplacian, rather than an ordinary $k^2$
approximation. We iterate until the residual is smaller than $10^{-9}$. Rather than solving each $(\sigma,h)$ point from scratch, we scan the parameter space by continuation~\cite{allgower1990numerical}, using the converged
solution at a nearby point as the initial guess for the next.

\section{Validation of the numerical saddle}
\label{sec:validation}

Every reported solution satisfies $\|\vF\|<10^{-9}$, where $\vF$ is the
residual of Eq.~\eqref{eq:eom}. However, a small residual alone does
not guarantee that the resulting action is accurate, so we validate each
point independently using the anisotropic Derrick identities derived in the
main text (Eqs.~(4)--(5)),
\begin{equation}
  \frac{T_\tau}{K_x}=\frac{\sigma}{2},\qquad
  \frac{U}{K_x}=\frac{\sigma-2}{2},\qquad
  B=\sigma K_x=2T_\tau,
  \label{eq:virial}
\end{equation}
which hold at the exact saddle for any tilt $h$ and are therefore independent
of the residual norm. We compare the measured ratios directly against these
predictions. Figure~\ref{fig:S1} shows the deviations $T_\tau/K_x-\sigma/2$ and
$U/K_x-(\sigma-2)/2$ across the full dataset ($\sigma\in[0.7,2]$, all
converged $h$): both remain below $2\%$. We also notice that points with smaller $\sigma$ satisfy the Derrick identities more accurately.

\begin{figure}[t]
  \centering
  \includegraphics[width=.9\linewidth]{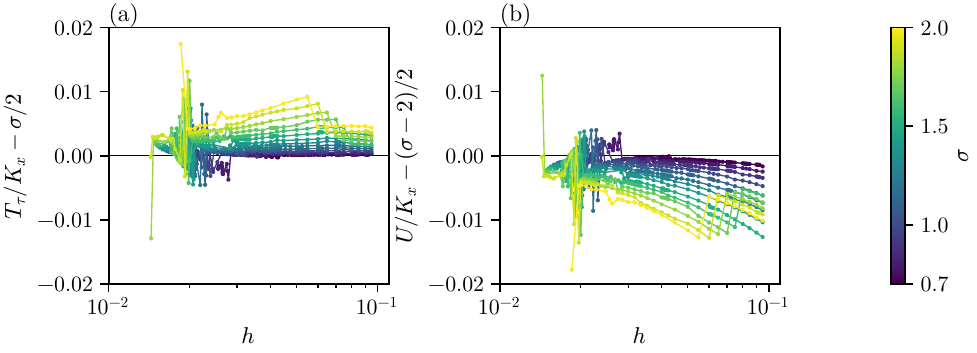}
  \caption{Validation of the numerical saddle using the anisotropic Derrick
  identities, main-text Eqs.~(4)--(5). (a) Deviation $T_\tau/K_x-\sigma/2$.
  (b) Deviation $U/K_x-(\sigma-2)/2$. Both vanish at the exact saddle for
  any tilt $h$; here they remain below $2\%$ throughout, with points
  colored by $\sigma$.}
  \label{fig:S1}
\end{figure}

As a final check, we compare with an independent pseudospectral solver of
the $(1+1)$-dimensional local Coleman problem at $h=0.1$, by setting $\sigma=2$, which formally reproduces the same limit. The two methods agree to within a few tenths of a percent.

\section{Beyond a single power law: the regime-dependent scaling fit}
\label{sec:scaling}

The main text fits the numerical action to the regime-dependent thin-wall form
\begin{equation}
  B_{\rm fit}(h) = C_1 h^{-\vartheta} + C_2 f_\sigma(h) + C_{\rm reg},
  \label{eq:fitform}
\end{equation}
The leading exponent is predicted to be $\vartheta=1/\sigma$ for $0<\sigma<1$ and $\vartheta=1$ for $1<\sigma<2$. In the fits reported in the main text, however, $\vartheta$ is left free, while the exponent of the first subleading term is fixed to its thin-wall value, by using $f_\sigma(h)=h^{-1}$ for $0<\sigma<1$ and $f_\sigma(h)=h^{\sigma-2}$ for $1<\sigma<2$. The constant $C_{\rm reg}$ accounts for regular corrections arising from the analytic dependence of the field jump, wall tensions, and vacuum splitting on $h$. The marginal case $\sigma=1$ is fitted separately to the logarithmic form given in the main text.

For each $\sigma$, we compare Eq.~\eqref{eq:fitform} with a pure leading-power fit, 
\begin{equation} 
    B_{\rm pure}(h)=a h^{-\vartheta_{\rm pure}}. 
    \label{eq:pure_power_fit} 
\end{equation} 
The pure power law is fitted by linear regression in logarithmic variables, whereas Eq.~\eqref{eq:fitform}, for which $\vartheta$ is free, is fitted by nonlinear least squares. All fits use the same set of converged data points.

We quantify the fit quality using the relative RMS misfit
\begin{equation}
\delta_{\rm RMS}
=
\left[
\frac{\sum_i \left(B_i-B_{\rm fit}(h_i)\right)^2}
{\sum_i B_i^2}
\right]^{1/2},
\end{equation}
where $B_i$ are the numerical actions. Figure~\ref{fig:S2}(a) shows the
relative RMS misfit of both models versus $\sigma$. The regime-dependent form
improves on the pure power law by roughly two orders of magnitude uniformly
across the dataset. 

Figure~\ref{fig:S2}(b) shows why a pure power law is
misleading: the fitted exponent $\vartheta$ is systematically steeper than the leading exponent predicted by main-text Eq.~(8).  The  bias
is largest at small $\sigma$, where the leading and subleading terms remain closest in size over the accessible range of $h$.

\begin{figure}
  \centering
  \includegraphics[width=.8\linewidth]{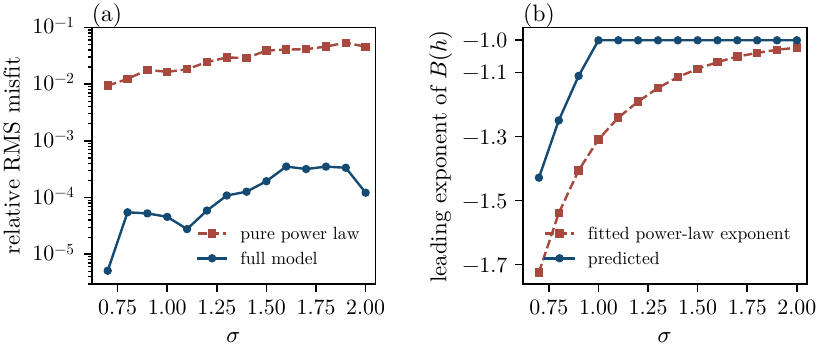}
  \caption{Direct comparison between a pure power-law fit and the
  regime-dependent form of main-text Eq.~(9). (a) Relative RMS misfit versus
  $\sigma$: the full model improves on the pure power law by about two orders
  of magnitude throughout. (b) Fitted leading exponent of $B(h)$: the pure
  power-law fit (squares) is systematically steeper than the predicted
  exponent $\vartheta$ (circles, main-text Eq.~(8)), most severely at small
  $\sigma$.}
  \label{fig:S2}
\end{figure}

\FloatBarrier
\section{Bubble anisotropy across the crossover}
\label{sec:profiles}

We finally show representative bounce profiles across the crossover from the
IR-dominated long-range regime to the UV-dominated, nearly local regime.
Figure~\ref{fig:S3} shows the shifted field $\varphi(x,\tau)$ at fixed tilt
$h=0.04$ for four values of $\sigma$. 
For $\sigma=0.7$ the bubble is strongly elongated along the spatial direction,
with $R_x\gg R_\tau$, reflecting the IR-dominated regime and the long spatial
tail generated by the fractional kernel. As $\sigma$ increases, the spatial
kernel becomes progressively more local and the profiles become more compact
and nearly isotropic. The right panel quantifies this trend through the
anisotropy ratio $R_x/R_\tau$: the solid gray curve is obtained from all
available $h=0.04$ solutions, while the colored markers correspond to the four
profiles shown on the left.
\begin{figure}
  \centering
  \includegraphics[width=\linewidth]{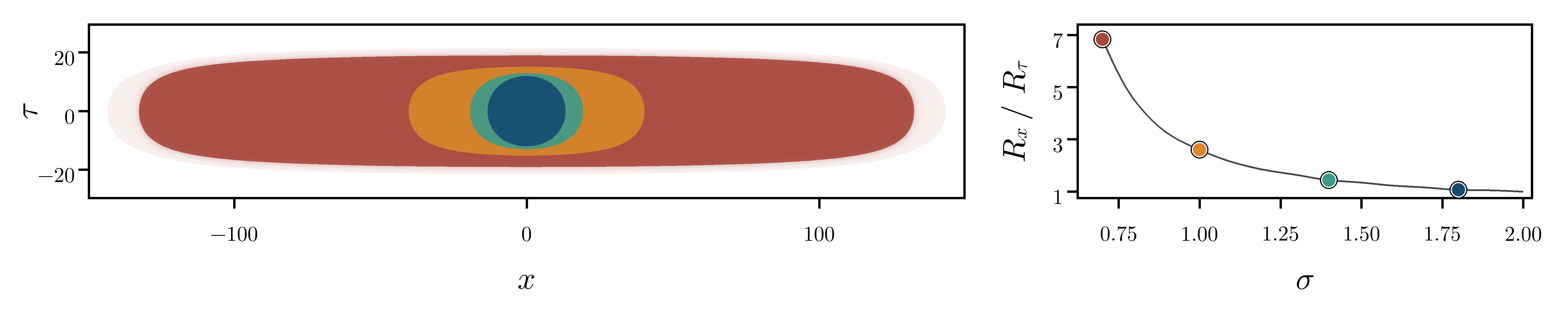}
  \caption{
Nested bounce profiles and anisotropy at fixed $h=0.04$.
Left: field profiles for $\sigma=0.7,1.0,1.4$, and $1.8$, overlaid on common
equal $x$ and $\tau$ scales. Right: aspect ratio $R_x/R_\tau$ versus $\sigma$.
The solid gray curve uses all available $h=0.04$ solutions, while the colored
markers indicate the four profiles shown on the left. The monotonic decrease
toward $R_x/R_\tau\simeq1$ shows the approach to isotropy as $\sigma$ increases.
}
  \label{fig:S3}
\end{figure}

Figure~\ref{fig:S4} illustrates the different asymptotic decays in space and
imaginary time. We use the $\sigma=0.7$ profile, where the spatial tail is
longest and the dynamic range is widest. The spatial cut $\varphi(x,0)$
displays the algebraic decay
$
\varphi(x,0)\sim |x|^{-(1+\sigma)}
$, over an intermediate range of $x$. At smaller
distances, deviations reflect the crossover from the bubble wall to the
asymptotic tail. Near the edge of the computational domain, the decay is
affected by the finite spatial box.

By contrast, the temporal cut $\varphi(0,\tau)$ decays exponentially. This is
consistent with the locality of the temporal operator, $\partial_\tau^2$,
and contrasts with the algebraic spatial decay generated by the fractional
Laplacian.

\begin{figure}[t]
  \centering
  \includegraphics[width=.75\linewidth]{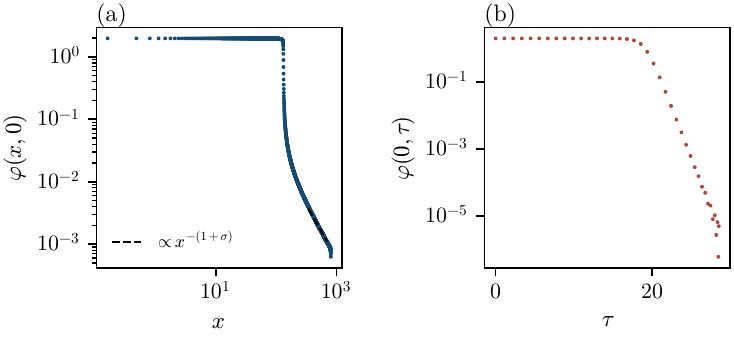}
  \caption{
Anisotropic bounce profile for $\sigma=0.7$, $h=0.04$.
(a) Spatial cut $\varphi(x,0)$, showing the algebraic tail.
(b) Temporal cut $\varphi(0,\tau)$, showing exponential decay. The contrast between the two panels reflects the fractional spatial
operator and the local temporal operator.
}
  \label{fig:S4}
\end{figure}

\end{document}